\documentclass[12pt,journal,twocolumn]{IEEEtran}
\usepackage{times,epsfig,amsbsy,amssymb,float,color}
\usepackage[cmex10]{amsmath}
\usepackage{mdwmath, mathrsfs, bbm}
\usepackage{mdwtab}
\usepackage{amsfonts}
\usepackage{subfigure}
\usepackage{multicol}
\usepackage{amsmath,amssymb}
\usepackage{cases}
\usepackage{booktabs}
\usepackage{algorithm}
\usepackage{stfloats}
\usepackage{color}
\usepackage{algorithmicx}
\usepackage{algpseudocode}
\usepackage{mathtools}
\usepackage{authblk}
\algdef{SE}[DOWHILE]{Do}{doWhile}{\algorithmicdo}[1]{\algorithmicwhile\ #1}
\DeclareMathOperator*{\argmin}{arg\,min}
\newtheorem{theorem}{Theorem}

\newtheorem{remark}{Remark}
\IEEEoverridecommandlockouts

\usepackage[numbers,sort&compress]{natbib}

\usepackage{flushend}

\usepackage{stackengine}
\usepackage{scalerel}
\def\clipeq{\!\mathrm{=}\!}
\def\Lla{\Longleftarrow\!\!\!\!}
\def\Lra{\!\!\!\!\Longrightarrow}
\newcommand\xleftrightarrows[3][0]{%
  \mathrel{%
  \ifthenelse{\equal{#1}{0}}%
  {\stackunder[2pt]{\stackon[3pt]{$\Lla\Lra$}%
     {$\scriptstyle#2$}}{$\scriptstyle#3$}}%
  {\stackunder[2pt]{\stackon[3pt]{$\Lla\hstretch{#1}{\clipeq}\Lra$}%
     {$\scriptstyle#2$}}{$\scriptstyle#3$}}%
  }%
}

\input epsf
\title{\huge Physical Layer Network Coding in Network MIMO: A New Design for 5G and Beyond}
\author{Tong~Peng, Yi~Wang, Alister G. Burr and Mohammad Shikh-Bahaei \vspace{-1em}

\thanks{T. Peng was with the Department of Electronics, University of York, and now is with the Centre for Telecommunications Research, Department of Informatics, King's college London, UK (e-mail: tong.peng@york.ac.uk, tong.peng@kcl.ac.uk). Y. Wang and A. G. Burr are with the Department of Electronics, University of York, UK	
(e-mails: yi.wang@york.ac.uk, alister.burr@york.ac.uk). M. Shikh-Bahaei is with the Centre for Telecommunications Research, Department of Informatics, King's college London, UK (e-mail: m.sbahaei@kcl.ac.uk).

This research is funded by EPSRC NetCoM project EP/K040006/1 and partially by EPSRC IoSIRE project EP/P022723/1.}
}

\begin{document}
\maketitle\pagestyle{empty}
\begin{abstract} Physical layer network coding (PNC) has been studied to serve wireless network MIMO systems with much lower backhaul load than approaches such as Cloud Radio Access Network (Cloud-RAN) and coordinated multipoint (CoMP). In this paper, we present a design guideline of engineering applicable PNC to fulfil the request of high user densities in 5G wireless RAN infrastructure. Unlike compute-and-forward and PNC design criteria for two-way relay channels, the proposed guideline is designed for uplink of network MIMO (N-MIMO) systems. We show that the proposed design criteria guarantee that 1) the whole system operates over binary system; 2) the PNC functions utilised at each access point overcome all singular fade states; 3) the destination can unambiguously recover all source messages while the overall backhaul load remains at the lowest level. We then develop a two-stage search algorithm to identify the optimum PNC mapping functions which greatly reduces the real-time computational complexity. The impact of estimated channel information and reduced number of singular fade states in different QAM modulation schemes is studied in this paper. In addition, a sub-optimal search method based on lookup table mechanism to achieve further reduced computational complexity with limited performance loss is presented. Numerical results show that the proposed schemes achieve low outage probability with reduced backhaul load.

\end{abstract}\vspace{-0.5em}

\vspace{-1em}

\section{Introduction}
\label{sec:introduction}

Over the past few years, network multiple-input multiple-output (N-MIMO) technique \cite{M.V.Clark} has received significant attention due to its flexibility, power and capacity advantage over the centralized architectures in fifth generation (5G) dense cellular networks. Multiple mobile terminals (MTs) may share the same radio resources and be served by the corresponding access points (APs) where the inter-cell interference can be effectively mitigated. This was also applied in the coordinated multipoint (CoMP) approach standardized in LTE-A \cite{LTE-A, R.Irmer}. The Cloud Radio Access Network (C-RAN) concept has been proposed in \cite{R.Irmer} with similar goals. However a potential issue in these approaches is a significant increased upload burden on the backhaul (referred to as fronthaul in C-RAN) network between APs and central processing unit (CPU) on the uplink, especially for wireless backhaul networks. Several methods have been studied in previous work in order to address this problem. In \cite{re-bkload0}, Wyner-Ziv compression is utilised to reduce the backhaul load. Iterative interference cancellation and compressive sensing algorithms are designed in \cite{re-bkload2}-\cite{nonidcran} as alternative solutions, but the total backhaul load remains typically several times the total user data rate. In \cite{Lattice_QTSun, DongF}, novel approaches have been designed based on physical layer network coding (PNC) that keep the total backhaul load equal to the total user data rate.

PNC is a scheme implemented at APs in which each AP attempts to infer and forward combinations of the signals over an algebraic field, where the signals are transmitted from multiple sources simultaneously and superimposed in the received constellation. An important property of PNC is that the APs decode the joint messages from multiple sources to a linear function over the algebraic field rather than decode each source symbol individually. On the other aspect, PNC is a multiple-message compressing technique that makes network throughput greatly improved and keeps the cardinality of relay outputs considerably reduced. Hence PNC is an appealing technique to serve wireless RAN infrastructure with high user density in 5G and beyond. 

Previous work on PNC mainly focused on a two-way relay channel (TWRC) scenario to easily double the network throughput without routine operations \cite{LPNC}-\cite{DongF2}. The original PNC was proposed and designed in a TWRC based on BPSK \cite{Zhang2006}. Although only BPSK was used, PNC contributes to a big idea and motivates many research outcomes thereafter, e.g. compute-and-forward (C$\&$F) which generalizes PNC of TWRC to multiuser relay networks by utilizing structured nested lattice codes \cite{CF_Nazer}, and lattice network coding \cite{FengLNC}. However, the lattice based network coding in construction A and D operates over a finite field and the coset size of the quotient lattices is typically not binary-based \cite{Zhang2006}. Then lattice codes have disadvantages for engineering applications as non-binary codes are required over large prime fields. 


In contrast to the previous work in PNC, we focus on designing the PNC approach with conventional $2^m$-ary digital modulation. When QAM modulation schemes are used at MTs, PNC has to solve the so-called singular fading problem which is typically unavoidable at the multiple access phase under some circumstances at each AP. Failure to resolve such problem results in network performance degradation. Toshiaki \textit{et. al.} \cite{Popovski, Popovski2} proposes a scheme, namely the denoise-and-forward, which employs a non-linear $5$QAM PNC mapping to mitigate all singular fade states, and gives good performance. Other researches on this issue have worked on the design of linear functions over the integer finite field or ring, e.g. linear PNC (LPNC) \cite{ABurr} which can only be optimised for the $q$-ary PNC mapping where $q$ is a prime in $\mathbb{Z}^+$. All these approaches, however, do not operate over the binary systems, and hence cannot be readily applied in the current mobile communication networks. The work in \cite{DongF} provides a solution for implementing PNC in binary systems with low modulation orders only.

In this paper, we propose an adaptive PNC with reduced backhaul load and unambiguous decoding for QAM modulation schemes and the main contributions are listed as follows
\begin{enumerate}

\item A PNC design guideline for uplink scenarios is proposed, along with a search algorithm based on this guideline to find the optimal coefficient mapping matrices, such that a) the global matrix, formed by the coefficient matrices from each AP, guarantees all source symbols to be decoded at CPU; b) the matrices stored at each AP can resolve all singular fade states; c) the number of coefficient matrices stored at each AP is minimised; d) the proposed algorithm is generalised to QAM modulation schemes of different orders; e) the proposed algorithm can be applied in N-MIMO systems with multiple MTs and APs. 

\item The whole scheme operates over binary systems with multiple MTs and APs. As discussed earlier about advantages of coping with the singular fading problem in multiple access stage, PNC plays as a reliable role not only in TWRC but also in RAN to serve multiple MTs. In this paper, we investigate the design criteria of engineering applicable PNC over binary systems for an uplink scenario of 5G N-MIMO system and discuss how to address the singular fading problem with multiple MTs.

\item A regulated PNC approach which fulfils the low latency demand in practical networks is also presented in this paper. The regulated approach is developed based on the original search algorithm, and the lookup table mechanism is adopted to achieve low latency. In such approach, all the optimal coefficient mapping matrices to resolve different singular fade state combinations are stored at the APs and the CPU with a corresponding table with their indexes. Instead of search among the matrix candidates, the optimal matrix selection algorithm is replaced by looking up the index table. The drawback of this approach is discussed in this paper and we provide a solution to overcome the problem. 

\item The impact of estimated channel information to optimal matrix selection as well as the effect of the reduced number of singular fade states to performance degradation are studied. The proposed PNC mapping selection algorithm depends on the accuracy of channel information at each AP, thus we tested how the estimated channel affects the proposed algorithm. Utilisation of less number of singular fade states contributes to reduced calculation complexity, but performance degradation is observed. We discuss these issues in this paper and give potential resolutions.


%

\end{enumerate}

The rest of this paper is organised as follows. The introduction of N-MIMO systems and definitions of PNC design criteria are given in Section II and III, respectively. The proposed binary matrix adaptive selection algorithm is derived in Section IV, followed by the discussion of methods to reduce the computational complexity of the proposed algorithm in Section V. Numerical results are given in Section VI and finally the conclusions are drawn in Section VII.

\section{System Model}
\begin{figure}[t]
  \centering
\begin{minipage}[t]{1\linewidth}
\centering
  \includegraphics[width=1\textwidth]{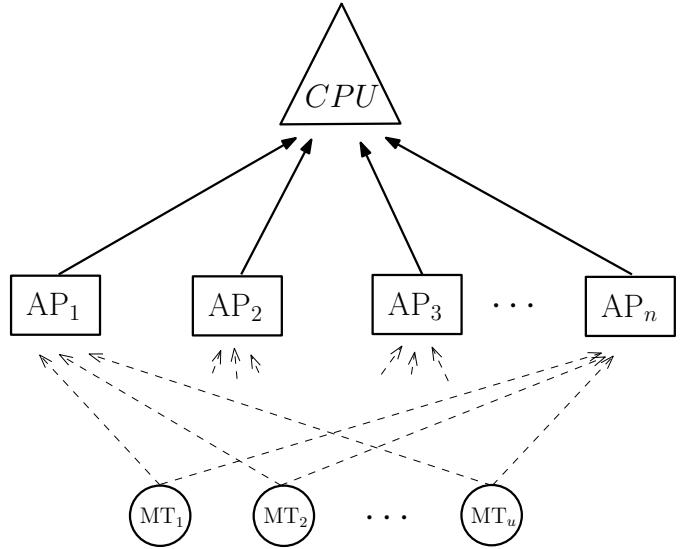}
\end{minipage}
\caption{\small The uplink system diagram.} \label{fig:system}
\end{figure}

A two-stage uplink model of N-MIMO system is illustrated in Fig. \ref{fig:system}. We assume MTs and APs are all equipped with a single antenna for simplicity. At the first stage, $u$ MTs transmit symbols to $n$ APs during the same period, which refers to a multi-access stage. We have studied the impact of synchronisation errors in \cite{Acs} so in this paper we assume the synchronisation is perfect for simplicity. Each AP receives data from all MTs and then infers and forwards a linear combination (which is referred to as the network coded symbols (NCS) in this paper) of the entire messages over a finite field or ring. The second stage is called backhaul stage where $n$ APs forward the NCSs to CPU via a lossless but capacity limited `bit-pipe'. In this paper, the links in multi-access stage are modelled as wireless links in order to fulfil the request of 5G systems; while the backhaul links may be wireless or deployed on wireline. The techniques presented in this paper are in particular suitable for wireless backhaul which is normally more cost-effective.

Each MT employs a $2^m$-ary digital modulation scheme where $m$ denotes the modulation order. Let $\mathscr{M}:\mathbb{F}_{2^m} \longrightarrow \Omega$ denotes a one-to-one mapping function, where $\Omega$ is the set of all possible complex constellation points. Hence the messages $\mathbf{w}_{\ell}\in \mathbb{F}_{2^m}$ at the $\ell^{\mathrm{th}}$ MT can be mapped to the complex symbol $s_{\ell} = \mathscr{M}(\mathbf{w}_{\ell})$, where $\mathbf{w}_{\ell} = [w_{\ell}^{(1)},\cdots,w_{\ell}^{(m)}]$ is an $m$-tuple with each element $w_{\ell}^{(i)}\in\mathbb{F}_2$.

The link between all MTs and the $j^{\mathrm{th}}$ AP forms a multiple access channel (MAC), where the $j^{\mathrm{th}}$ AP observes the noisy, faded and superimposed signals at a certain time slot, mathematically given by
\begin{align}
y_{j} = \sum_{\ell=1}^u h_{j,\ell}s_{\ell} + z_j, \label{equ:channel}
\end{align}
where $z_j$ denotes the additive complex Gaussian noise with zero mean and variance $\sigma^2$, and $h_{j,\ell}$ represents the channel fading coefficient between the ${\ell}^{\mathrm{th}}$ MT and the $j^{\mathrm{th}}$ AP, which is a random variable with Rayleigh distribution. 

\section{Design Criteria}
\label{sec:UD}

Before presenting the proposed PNC design criteria, we list three main constraints for a PNC to be engineering applicable:
\begin{enumerate}
\item The PNC decoding must operate over $\mathbb{F}_2$; thus, the NCS need to be binary-based.
\item The PNC mapping function must be well designed such that all singular fade states can be resolved.
\item The PNC mapping functions must ensure that CPU can unambiguously recover all source messages. 
\end{enumerate}

We give details of the proposed design criteria for PNC in N-MIMO systems in this section, which relaxes all three aforementioned constraints.

\subsection{Engineering Applicable PNC Function}
We are primarily concerned with the MAC phase between $u$ MTs and the $j^{\mathrm{th}}$ AP in the design of the PNC mapping function. Instead of using PNC approaches performing linear combinations on symbol level, such as \cite{LPNC}, we design a method to encode PNC directly at the bit level which allows the APs to operate over a binary field for industrial application.

\textit{Definition 1:} The bit-level linear network coding function of the $j^{\mathrm{th}}$ AP for $u$ MTs is defined as
\begin{align}
\mathscr{N}_j: (\mathbf{M}_j,\mathbf{w}) \longrightarrow \mathbf{x}_j,
\end{align}
and mathematically expressed as 
\begin{align}\label{eq:NCSt}
\mathbf{x}_j = \mathscr{N}_j(\mathbf{M}_j,\mathbf{w}) = \mathbf{M}_j \otimes \mathbf{w},
\end{align}
where $\mathbf{w}\triangleq [\mathbf{w}_1,\cdots,\mathbf{w}_u]^T$ denotes an $mu \times 1$ joint message set with $\mathbf{w}\in\mathbb{F}_2^{mu \times 1}$, and each $\mathbf{w}_\ell$ stands for a $1 \times m$ binary data vector at the $\ell^{\mathrm{th}}$ MT. $\mathbf{M}_j$ denotes a matrix with $\mathbb{F}_2^{t \times mu}$, where $t$ stands for the size of network coded vector at the $j^{\mathrm{th}}$ AP and $t \geq m$, and $\otimes$ denotes the multiplication over $\mathbb{F}_2$. $\mathbf{x}_j\in\mathbb{F}_2^{t \times 1}$ is called the network coded vector (NCV) which consists of all $t$ linear network coded bits
\begin{align}
\mathbf{x}_j = [x_j^{(1)},x_j^{(2)},\cdots,x_j^{(t)}]^T.
\end{align} $\square$

It is obvious that each coded bit $x_j^{(i)}$ is indeed a linear combination of all source bits over $\mathbb{F}_2$, thus,
\begin{align}
x_j^{(i)} = M_j^{(i,1)}\otimes w_{1}^{(1)}\oplus\cdots\oplus M_j^{(i,um)}\otimes w_{u}^{(m)},
\end{align}
where $\oplus$ denotes the addition operation over $\mathbb{F}_2$, and $M_j^{(i,1)}$ denotes the entry at the $i^{\mathrm{th}}$ row and the $1^{\mathrm{st}}$ column of $\mathbf{M}_j$.

\textit{Definition 2:} We define the constellation set which contains all possible superimposed symbols at the $j^{\mathrm{th}}$ AP over a given channel coefficient vector $\mathbf{h}_j\triangleq[h_{j,1},\cdots,h_{j,u}]$ as $\mathbf{s}_{j,\bigtriangleup}\triangleq[s^{(1)}_{j,\bigtriangleup},\cdots,s^{(2^{mu})}_{j,\bigtriangleup}]$, where
\begin{align}
s^{(\tau)}_{j,\bigtriangleup} = \sum_{\ell=1}^u h_{j,\ell}s_{\ell}, ~~\forall s_{\ell}\in\Omega,
~\tau = 1,2,\cdots,2^{mu}. \notag \label{equ:SIC.Definition}
\end{align}

\begin{theorem}
For the MAC link between $u$ MTs and the $j^{\mathrm{th}}$ AP, there exists a surjective function
\begin{align}
\mathrm{\Theta}: \mathbf{s}_{j,\bigtriangleup} \longrightarrow \mathbf{x}_j,
\end{align}
\end{theorem}
when the size of NCV $t<mu$.

\begin{IEEEproof}
Since $\mathscr{M}$ is a bijective function, we have the following relationship
\begin{align}
\mathbf{x}_j  \xLeftarrow{\mathscr{N}_j} \mathbf{w}\xleftrightarrows{\mathscr{M}}{\mathscr{M}^{-1}} \mathbf{s},
\end{align}
where $\xLeftarrow{}$ and $\xleftrightarrows{}{}$ represent surjective and bijective relationships, respectively. $\mathbf{s}\triangleq [s_1,\cdots,s_u]=[\mathscr{M}(\mathbf{w}_1), \cdots, \mathscr{M}(\mathbf{w}_u)]$ stands for the set that contains the modulated symbols at all MTs. Following (\ref{equ:SIC.Definition}), for each element in $\mathbf{s}$, there exists a superimposed constellation point $s_{j,\bigtriangleup}$ at a given channel coefficient vector $\mathbf{h}_j$, and this proves Theorem 1.
\end{IEEEproof}

We call $\mathrm{\Theta}$ the PNC mapping function which maps a superimposed constellation point to an NCV and plays the key role in PNC encoding, where this PNC encoding performs estimation of the possible NCV outcomes $\mathbf{x}_j$ for the $j^{\mathrm{th}}$ AP, based on the received signals $y$. Let $\mathbf{X}_j$ denote the vector-based random variable with its realization $\mathbf{x}_j$. The \textit{a posteriori} probability of the event $\mathbf{X}_j = \mathbf{x}_j$ conditioned on the MAC outputs $Y_j=y_j$ is
\begin{align}
&\mathrm{Pr}(\mathbf{X}_j = \mathbf{x}_j|y_j,\mathbf{h}_j) \notag \\
=&\frac{\mathrm{Pr}(Y_j|\mathbf{X}_j=\mathbf{x}_j,\mathbf{h}_j)\mathrm{Pr}(\mathbf{X}_j=\mathbf{x}_j)}{\mathrm{Pr}(Y_j=y_j)}  \notag \\
=&\frac{\sum\limits_{\forall\mathbf{w}:\mathscr{N}_j(\mathbf{M}_j,\mathbf{w})=\mathbf{x}_j}\mathrm{Pr}(Y_j|\mathbf{w},\mathbf{h}_j)\mathrm{Pr}(\mathbf{w})}{\mathrm{Pr}(Y_j=y_j)} \notag \\
=&\frac{\sum\limits_{\forall\mathbf{s}:\Theta(\mathbf{s}_{j,\bigtriangleup})=\mathbf{x}_j}\mathrm{Pr}(Y_j|\mathbf{S}_{j,\bigtriangleup}=\mathbf{s}_{j,\bigtriangleup})\mathrm{Pr}(\mathbf{S}=\mathbf{s}) }
{\mathrm{Pr}(Y_j=y_j)}. \label{equ:PostProb.bg}
\end{align}
The conditional probability density function is given by 
\begin{align}
\mathrm{Pr}(Y_j|\mathbf{S}_{j,\bigtriangleup}=\mathbf{s}_{j,\bigtriangleup}) = \frac{1}{\sqrt{2\pi\sigma^2}}\mathrm{exp}\left(-\frac{|y_j-\mathbf{s}_{j,\bigtriangleup}|^2}{2\sigma^2}		\right).
\end{align}
The \textit{a posteriori} L-value $L_{\mathbf{x}_j}$ for the event $\mathbf{X}_j = \mathbf{x}_j$ is
\begin{align}\label{equ:PostProb.ed}
L_{\mathbf{x}_j} = \log\left(\frac{\sum\limits_{\forall\mathbf{s}:\Theta(\mathbf{s}_{j,\bigtriangleup})=\mathbf{x}_j}\mathrm{Pr}(Y_j|\mathbf{S}_{j,\bigtriangleup}=\mathbf{s}_{j,\bigtriangleup})\mathrm{Pr}(\mathbf{S}=\mathbf{s}) }{\sum\limits_{\forall\mathbf{s}:\Theta(\mathbf{s}_{j,\bigtriangleup})=\mathbf{0}}\mathrm{Pr}(Y_j|\mathbf{S}_{j,\bigtriangleup}=\mathbf{s}_{j,\bigtriangleup})\mathrm{Pr}(\mathbf{S}=\mathbf{s}) }	\right),
\end{align}
where $\mathbf{0}$ is a length-$t$ all-zero vector over $\mathbb{F}_2^{t \times 1}$.

\subsection{Resolving the Singular Fading}
We have set up the PNC mapping approach in binary systems, which establishes the fundamental PNC system structure available for practical engineering application. The next upcoming problem lies in how to resolve the singular fading in the multiple access stage. In this section, we demonstrate that the PNC mapping function $\mathrm{\Theta}$ proposed above is capable of resolving all singular fade states with a simple design approach. We first define the singular fade states as follows

\textit{Definition 3:} The singular fade state (SFS) at the $j^{\mathrm{th}}$ AP is defined as the channel fading coefficients $\mathbf{h}_j$ which makes $s_{j,\bigtriangleup}^{(\tau)} = s_{j,\bigtriangleup}^{(\tau^{\prime})}$ when $\tau\neq\tau^{\prime}$. $\square$

In other words, for a given channel coefficient vector $\mathbf{h}_j$, if two or more elements in the set $\mathbf{s}_{j,\bigtriangleup}$ are the same, $\mathbf{h}_j$ is an SFS. Normally, singular fading is unavoidable at MAC stage, and multiuser detection is in principle infeasible if the $j^{\mathrm{th}}$ AP expects to decode all source messages. PNC is capable to overcome SFS problem when the coincident superimposed constellation points are well labelled by the NCV $\mathbf{x}_j$, which helps CPU to recover all source messages.

\textit{Definition 4:} If a set of constellation points received at APs are mapped to the same NCV, we call this set a \emph{cluster}, denoted as
\begin{align}
\mathbf{c}^{(\tau)} \triangleq [s^{(\tau_1)}_{j,\bigtriangleup}, s^{(\tau_{2})}_{j,\bigtriangleup}, \cdots],
\end{align}
where $s^{(\tau_i)}_{j,\bigtriangleup}$ denotes the $i^{\mathrm{th}}$ cluster members. In a singular fading, if the values of cluster members are the same then this cluster is called a \emph{clash}. $\square$

\textit{Definition 5:} Given two clusters $\mathbf{c}^{(\tau)}$ and $\mathbf{c}^{(\tau^\prime)}$ that the constellation points in which are mapped to different NCVs, then the minimum inter-cluster distance, also known as the minimum distance between these two different NCVs, is defined as
\begin{align}
d_{\mathrm{min}} = &\min_{\Theta(s_{j,\bigtriangleup}^{(\tau_i)})\neq \Theta(s_{j,\bigtriangleup}^{(\tau^{\prime}_k)}) }  |s_{j,\bigtriangleup}^{(\tau_i)} - s_{j,\bigtriangleup}^{(\tau^{\prime}_k)}|^2, \\
\forall s^{(\tau_i)}_{j,\bigtriangleup} \in \mathbf{c}^{(\tau)}, &~\forall s^{(\tau^\prime_k)}_{j,\bigtriangleup} \in \mathbf{c}^{(\tau^\prime)}, ~i=1,2,..., ~k=1,2,... \notag \label{dmint}
\end{align}



\begin{theorem}
The PNC mapping function $\mathrm{\Theta}$ cannot resolve singular fading if the minimum inter-cluster distance $d_{\mathrm{min}} = 0$.
\end{theorem}
\begin{IEEEproof}
When $d_{\mathrm{min}} = 0$, the posterior probability of some outcomes of $\mathbf{X}_j$ will be very similar (in terms of (\ref{equ:PostProb.bg})). This definitely introduces the ambiguities in estimating the real NCV $\mathbf{x}_j$, especially when a superimposed constellation point labelled by one NCV is close to another point that is labelled by another NCV. Hence the singular PNC mapping function is in principle not capable of decoding the NCV reliably.
\end{IEEEproof}

Normally the dimension of NCV $\mathbf{x}_j$ is $t \times 1$ at the $j^{\mathrm{th}}$ AP, $m \leq t \leq mu$. When the number of source increases (a large MAC), the singular fading problem becomes more severe and the method of SFS values calculation is different. However, by simply increasing the dimension $t$ of NCV (thus, increasing the number of rows of $\mathbf{M}_j$), there definitely exists non-singular PNC function which is capable of resolving a kind of SFS.

\begin{remark} \label{remark:t}
We can obtain non-singular PNC mapping function $\mathrm{\Theta}_j$ for the $j^{\mathrm{th}}$ AP if the cardinality $t$ of the PNC encoding outcomes are determined in terms of the following criterion
\begin{align}
t = \argmin_{m\leq t<mu} \left\{d_{\mathrm{min}} - d_{\alpha} \geq 0\right\},
\end{align}
where $d_{\alpha}>0$ is a distance threshold.
\end{remark}

Remark \ref{remark:t} reveals the second design criterion for PNC mapping function $\mathrm{\Theta}_j$ over a $u$-MT and $2^m$-ary digital modulation MAC, which guarantees the reliable PNC encoding with the minimum possible cardinality expansion.



\subsection{Algebraic Work for Unambiguous Decodability}
We have set up two design guidelines of the engineering applicable PNC approach for uplink scenarios. The next criterion is that the CPU can guarantee all source messages to be unambiguously recovered. We need to carefully design each $\mathbf{M}_j$, $j=1,2,\cdots,n$, so that $\mathbf{M} = [\mathbf{M}_1,\cdots,\mathbf{M}_n]^T$ includes a number of row coefficients which forms the following theorem:
\begin{theorem} \label{theorem:full.rank}
Assume $\mathbf{M}=M_{n\times n}(R)$, where the coefficients are from a commutative ring $R$. Source messages are drawn from a subset of $R$ and all source messages can be unambiguously decoded at the destination if and only if the determinant of the transfer matrix is a unit in $R$,
\begin{align}
\mathrm{det}(\mathbf{M}) = \mathcal{U}(R). \label{equ:det.unit}
\end{align}
\end{theorem}

\begin{IEEEproof} We first prove that (\ref{equ:det.unit}) gives the sufficient and necessary conditions that make a matrix $\mathbf{B}$ invertible in N-MIMO networks. Suppose $\mathbf{B}$ is invertible, then there exists a matrix $\mathbf{C}\in M_{n\times n}(R)$ such that $\mathbf{B}\mathbf{C} = \mathbf{C}\mathbf{B} = \mathbf{I}_n$. This implies $1 = \mathrm{det}(\mathbf{I}_n) = \mathrm{det}(\mathbf{B}\mathbf{C}) = \mathrm{det}(\mathbf{B})\mathrm{det}(\mathbf{C})$. According to the definition of a unit, we say $\mathrm{det}(\mathbf{B})\in U(R)$.

We know $\mathbf{B}\cdot \mathrm{adj}(\mathbf{B}) = \mathrm{adj}(\mathbf{B})\cdot \mathbf{B} = \mathrm{det}(\mathbf{B})\mathbf{I}_n$. If $\mathrm{det}(\mathbf{B})\in U(R)$, we have
\begin{align}
\mathbf{B} \cdot (\mathrm{det}(\mathbf{B})^{-1}\mathrm{adj}(\mathbf{B}) ) &=   (\mathrm{det}(\mathbf{B})^{-1}\mathrm{adj}(\mathbf{B}) ) \mathbf{B}  \notag\\
&= \mathrm{det}(\mathbf{B})^{-1} \mathrm{det}(\mathbf{B}) = \mathrm{I}_n.
\end{align}
Hence, $\mathbf{C} = (\mathrm{det}(\mathbf{B})^{-1}\mathrm{adj}(\mathbf{B}) )$ is the inverse of $\mathbf{B}$ since $\mathbf{B}\mathbf{C} = \mathbf{C}\mathbf{B} = \mathbf{I}_n$.

If $\mathbf{B}$ is invertible, then its inverse $\mathbf{B}^{-1}$ is uniquely determined. Assuming $\mathbf{B}$ has two inverses, say, $\mathbf{C}$ and $\mathbf{C}^{\prime}$, then
\begin{align}
\mathbf{B}\cdot \mathbf{C} &= \mathbf{C}\cdot \mathbf{B} = \mathbf{I}_n, \\
\mathbf{B}\cdot \mathbf{C}^{\prime} &= \mathbf{C}^{\prime}\cdot \mathbf{B} = \mathbf{I}_n,
\end{align}
hence we have
\begin{align}
\mathbf{C} = \mathbf{C}\cdot \mathbf{I}_n = \mathbf{C}\cdot \mathbf{B}\cdot \mathbf{C}^{\prime} = \mathbf{I}_n\cdot \mathbf{C}^{\prime} = \mathbf{C}^{\prime}.
\end{align}
It proves the uniqueness of the invertible matrix $\mathbf{B}$ over $R$.

Assume $\mathbf{a}\neq \mathbf{a^{\prime}}$, $\mathbf{B}\cdot \mathbf{a} = \mathbf{F}$, $\mathbf{B}\cdot \mathbf{a}^{\prime} = \mathbf{F^{\prime}}$, and $\mathbf{F} = \mathbf{F^{\prime}}$. This means
\begin{align}
\mathbf{a} = \mathbf{B}^{-1}\cdot \mathbf{F} = \mathbf{B}^{-1}\cdot \mathbf{F^{\prime}} = \mathbf{a^{\prime}}.
\end{align}
This contradicts $\mathbf{a}\neq \mathbf{a^{\prime}}$. Hence it ensures unambiguous decodability:
\begin{equation}
\mathbf{B}\cdot \mathbf{a} \neq \mathbf{B}\cdot \mathbf{a^{\prime}}, ~ \forall\mathbf{a}\neq \mathbf{a^{\prime}}.
\end{equation}
\end{IEEEproof}

\textit{Definition 6:} The ideal in $R$ generated by all $\nu\times\nu$ minors of $M_{m\times n}(R)$ is denoted by $I_{\nu}(M_{m\times n}(R))$, where $\nu=1,2,\cdots,r=\min\{m,n\}$. $\square$

A $\nu\times \nu$ minor of $M_{m\times n}(R)$ is the determinant of a $\nu\times\nu$ matrix obtained by deleting $m-\nu$ rows and $n-\nu$ columns. Hence there are $\binom{m}{\nu}\binom{n}{\nu}$ minors of size $\nu\times\nu$. $I_{\nu}(M_{m\times n}(R))$ is the ideal of $R$ generated by all these minors.

\textit{Design Criterion}: The destination is able to unambiguously decode $u$ source messages if:
\begin{enumerate}
\item $u\geq\max\left\{\nu\mid \mathrm{Ann}_R(I_{\nu}(\mathbf{M}_j)) = \langle 0 \rangle \right\}$, $\forall j=1,2,\cdots,n$,
\item $\mathbf{M}_j = \arg\max\limits_{\mathbf{M}_j} \left\{ I\left( \overrightarrow{Y}; \overrightarrow{F}_j \right)  \right\}$,
\end{enumerate}
where $\langle x\rangle$ denotes the ideal generated by $x$. 

Condition 1 can be proved as follows. According to Laplace's theorem, every $(\nu+1)\times (\nu+1)$ minor of $M_{m\times n}(R)$ must lie in $I_{\nu}(M_{m\times n}(R))$. This suggests an ascending chain of ideals in $R$:
\begin{align}
\langle 0 \rangle = I_{r+1}(\mathbf{M}_j)&\subseteq I_{r}(\mathbf{M}_j)\subseteq\cdots\subseteq I_{1}(\mathbf{M}_j)\notag\\
&\subseteq I_{0}(\mathbf{M}_j) =R. \label{equ:chain}
\end{align}
Computing the annihilator of each ideal in (\ref{equ:chain}) produces another ascending chain of ideals,
\begin{align}
\langle 0 \rangle &=\mathrm{Ann}_R(R)\subseteq\mathrm{Ann}_R(I_{1}(\mathbf{M}_j))\subseteq\cdots \notag\\
&\subseteq\mathrm{Ann}_R(I_{r}(\mathbf{M}_j))\subseteq\mathrm{Ann}_R(\langle 0 \rangle ) =R.
\end{align}
It is obvious that:
\begin{align}
&\mathrm{Ann}_R(I_{k}(\mathbf{M}_j))\neq \langle 0 \rangle \notag \\
\Rightarrow & \mathrm{Ann}_R(I_{k^\prime}(\mathbf{M}_j))\neq \langle 0 \rangle, ~~\forall k\leq k^\prime.
\end{align}

The maximum value of $\nu$ which satisfies $\mathrm{Ann}_R(I_{\nu}(\mathbf{M}_j))= \langle 0 \rangle$ guarantees that $I_{k}(\mathbf{M}_j)\in R$, $\forall k<\nu$. Hence we define the rank of $\mathbf{M}_j$ as $\mathrm{rk}(\mathbf{M}_j)=\max\left\{\nu\mid \mathrm{Ann}_R(I_{\nu}(\mathbf{M}_j)) = \langle 0 \rangle \right\}$. Suppose that $\mathbf{M}_k\in M_{m\times p}(R)$ and $\mathbf{M}_{k^\prime}\in M_{p\times n}(R)$, then $\mathrm{rk}(\mathbf{M}_k\mathbf{M}_{k^\prime})\leq \min\{\mathrm{rk}(\mathbf{M}_k),\mathrm{rk}(\mathbf{M}_{k^\prime})\}$, and we can easily prove that $0\leq \mathrm{rk}(M_{m\times n}(R))\leq \min\{m,n\}$. Thus, in order to guarantee there are at least $u$ unambiguous linear equations available at the CPU, $\mathrm{rk}(\mathbf{M}_j)$ must be at least $u$, $\forall j=1,2,\cdots,n$.

The special case of condition $1$ is that the entry of the coefficient matrix $\mathbf{M}_j\in M_{m\times n}(F)$ is from a finite field $F\in\mathbb{F}$. Then condition $1$ of the above \textit{Design Criterion} may be changed to \textquotedblleft the maximum number of linearly independent rows (or columns)\textquotedblright~ since $\mathrm{Ann}_R(I_{\nu}(\mathbf{M}_j))= \langle 0 \rangle$ if and only if $I_{\nu}(\mathbf{M}_j)\neq 0$. In other words, the largest $\nu$ such that the $\nu\times \nu$ minor of $\mathbf{M}_j$ is a non-zero divisor represents how many reliable linear combinations the $j^{\mathrm{th}}$ layer may produce. Hence condition $1$ is a strict definition which ensures unambiguous decodability of the $u$ sources. Condition $2$ ensures that the selected coefficient matrix maximises the mutual information of the particular layer, giving finally the maximum overall throughput.

\section{Binary Matrix Adaptive Selection Algorithm Design}
\label{sec:CombInfo}
According to the design criteria proposed in the previous section, we can summarise that given a QAM modulation scheme, the optimal binary PNC mapping function contains the following properties

\begin{enumerate}
\item it maximises the minimum distance between different NCVs ;
\item the composited global mapping matrix is invertible.
\end{enumerate}

In order to achieve these properties and applicability in practical N-MIMO systems, we propose a binary matrix adaptive selection (BMAS) algorithm based on the design criteria introduced in Section III. The BMAS algorithm is divided into two stages, one is called Off-line search, in which an exhaustive search is implemented among all $m \times mu$ binary mapping matrices to find a set of candidate matrices which resolves all SFSs with the above property 1); the other one is called On-line search, in which a selection from the candidate matrices in order to obtain the invertible mapping matrix according to property 2) is executed. The computational complexity of the proposed algorithms is mainly caused by the Off-line search, especially in higher order modulation schemes due to the increased number of SFSs and matrices, and this Off-line search algorithm only needs to be done once for each modulation scheme. The candidate mapping matrices found in Off-line search are stored at APs and CPU in order to implement the On-line search in real-time transmission.

\subsection{Off-Line Search Algorithm}
Define a set $\mathbf{W}_{joint} \triangleq [\mathbf{w}_{jo_1}, \mathbf{w}_{jo_2}, \cdots,\mathbf{w}_{jo_N}]$ which contains all possible binary joint message combinations with $N=2^{um}$, so that each $\mathbf{w}_{jo_i}$ in this set stands for a $1 \times mu$ binary joint message vector from $u$ MTs, for $i=1,2,\cdots,N$. By applying a modulation scheme $\mathscr{M}$ over each $\mathbf{w}_{jo_i}$ in $\mathbf{W}_{joint}$, a joint modulation set $\mathbf{S}_{joint} \triangleq [\mathbf{s}_{jo_1}, \mathbf{s}_{jo_2}, \cdots,\mathbf{s}_{jo_N}]^{T}$ is obtained, where $\mathbf{s}_{jo_i}=[s^{(jo_i)}_1 ~s^{(jo_i)}_2 \cdots ~s^{(jo_i)}_u]$ stands for the $i^\mathrm{th}$ combination of $u$ modulated symbols and $s^{(jo_i)}_\ell \in \Omega$ for $\ell = 1, \cdots, u$. The next step is to calculate the NCS $s^{(q)}_{n,\bigtriangleup}$ and its corresponding NCV $\mathbf{x}^{(q)}_{i,n}$ under all $L$ SFS circumstances, mathematically given by
\begin{align}\label{eq:NCScwd}
& ~~~s^{(q)}_{n,\bigtriangleup} = \mathbf{h_v}^{(q)}_{SFS}\mathbf{s}_{jo_n}^T,  
 ~\mathbf{x}^{(q)}_{i,n} = \mathbf{M}_i \otimes \mathbf{w}^{(q)}_{jo_n},  \\ 
& n=1,\cdots,N, ~q=1,\cdots,L, ~i=1,\cdots, N^2, \notag
\end{align}
where $\mathbf{h_v}^{(q)}_{SFS}$ denotes the channel coefficient vector causes SFS. Due to the property of SFS, the same $s^{(q)}_{n,\bigtriangleup}$ could be obtained with different $\mathbf{s}_{jo_n}$ sets in a clash. In that case, these joint symbol sets should be encoded to the same NCV according to the unambiguous decodability theorem. The next step is to store the mapping matrices which can resolve one SFS and also contains a high possibility to form an invertible global mapping matrix when combining with other selected mapping matrix candidates in other SFSs. Detailed description of the Off-line search is illustrated in Algorithms \ref{Alg:offline.p1} and \ref{Alg:offline.p2} in Appendix \ref{ap:offline}.

\subsection{On-Line Search Algorithm}
The proposed Off-line search is implemented before the transmission to reduce the number of mapping matrices utilised in the On-line search. In the real-time transmission, the proposed On-line search, which contains the same steps in Algorithm \ref{Alg:offline.p2} but with a much smaller value of $K$, is applied at the CPU to select the optimal mapping matrix for each AP.

When the optimal mapping matrix is selected, the indexes of the selected mapping matrices will be sent back to each AP through the backhaul channel and at each AP, an estimator calculates the conditional probability of each possible NCV given the optimal mapping function. The estimator returns the log-likelihood ratio (LLR) of each bit of $\mathbf{x}_j$ which is then applied to a soft decision decoder. Note that the LLR algorithm does not require to detect individual symbols transmitted from each MT but a linear combination of the binary messages. Finally the NCV at each AP will be forwarded to the CPU and the original data from all MTs can be recovered by multiplying the inverse of the global binary PNC mapping matrix.

\section{Analysis and Discussion}
\label{sec:PrinSFS}
In this section, we discuss how to apply the proposed BMAS algorithm to a general N-MIMO network with multiple MTs and APs, including utilisation of reduced number of SFSs and discussion of resolving the SFS problem with more than $2$ MTs. With a study of the properties of SFSs, a regulated On-line search algorithm based on lookup table mechanism with a small performance degradation is proposed in this section in order to fulfil the request of low latency in 5G RANs.

\subsection{Image SFSs and Principal SFSs}
Since an exhaustive search is carried out among all $t \times mu$ binary matrices in the proposed Off-line search algorithm, the computational complexity increases due to a large number of SFSs as well as an increased value of $um$ in higher order modulation schemes with large number of MTs. For example, in $2$-MT and $2$-AP case, the number of SFSs need to be resolved at each AP is $L=13$ in $4$QAM and $L=389$ in $16$QAM. Thus for the $4$QAM case, at least $13$ binary matrices with each size of $2 \times 4$ should be stored at each AP for On-line search. When $16$QAM scheme is employed at each MT, at least $389$ of $4 \times 8$ binary matrices need to be stored which results in a huge increased number of candidates in real-time computation.

In the proposed BMAS algorithm, we resolve this problem by keeping the number of useful SFSs minimum. According to our research of NCV calculation expressed in (\ref{eq:NCScwd}), we found some of different SFSs generate the same clashes which can be resolved by the same binary matrices. We then define such SFSs as image SFSs (iSFSs) and keep only one in the proposed search algorithm.




However, this problem still remains when higher order modulation schemes are employed at MTs. In addition, due to a larger constellation in a higher order QAM scheme, a few SFSs cannot be resolved by a binary mapping matrix. In order to address these problems, we focused on the occurrence probability of an SFS in higher modulation schemes and noticed that not all SFSs occur frequently, so that we can ignore those \textquotedblleft nonactive\textquotedblright ~SFSs with low appearance probabilities to minimise the number of mapping matrices utilised in the proposed On-line search. We define the SFSs with high appearance probabilities as principal SFSs (pSFSs) and a trade-off between the performance degradation and the number of pSFSs used in Off-line search is illustrated in the next section.

\subsection{Calculation of Singular Fade States}

We illustrate how to determine a singular fading for a QAM modulation scheme with a simple network first with $u=2$ MTs, and discuss the SFS calculation issue in a network with more than $2$ MTs later. Following Definition $3$, given a QAM modulation scheme, for singular fading we have $s_{j,\bigtriangleup}^{(\tau)} = \ s_{j,\bigtriangleup}^{(\tau^{\prime})}$ when $\tau\neq\tau^{\prime}$ in a constellation. Then mathematically, an SFS can be derived as
\begin{align}
& s_{1,\bigtriangleup}^{(\tau)} = s_{1,\bigtriangleup}^{(\tau^{\prime})},   
~\mathbf{h}\mathbf{s}^{(\tau)} = \mathbf{h}\mathbf{s}^{(\tau^\prime)}, ~\text{for}~ \mathbf{s}^{(\tau)} \neq \mathbf{s}^{(\tau^\prime)}, \notag \\
& h_{1,1}s^{(\tau)}_1+h_{1,2}s^{(\tau)}_2 = h_{1,1}s^{(\tau^{\prime})}_1+h_{1,2}s^{(\tau^{\prime})}_2, \notag   
\end{align}
where $h_{j,\ell}$ denotes the channel coefficient between the $\ell^{\mathrm{th}}$ MT and the $j^{\mathrm{th}}$ AP, and $\mathbf{s}^{(\tau)}\triangleq[s^{(\tau)}_1 s^{(\tau)}_2]$ refers to a joint symbol set which contains the modulated symbols at both MTs, and $\mathbf{s}^{(\tau)} \neq \mathbf{s}^{(\tau^\prime)}$ means at least one symbol is different in $\mathbf{s}^{(\tau)}$ and $\mathbf{s}^{(\tau^\prime)}$. Then we define $\mathbf{v}_{SFS}=[v^{(1)}_{SFS}, v^{(2)}_{SFS}, \cdots , v^{(L)}_{SFS}]$ as the set contains all unique value of SFSs and calculated by
\begin{align}\label{eq:SFSt}
v^{(q)}_{SFS} = \frac{h_{1,2}}{h_{1,1}} = \frac{s^{(\tau)}_1-s^{(\tau^{\prime})}_1}{s^{(\tau^{\prime})}_2-s^{(\tau)}_2}, ~~\forall s^{(\tau)}_l,s^{(\tau^{\prime})}_l \in \Omega. 
\end{align}
By substituting the QAM modulated symbols with all possible combinations to (\ref{eq:SFSt}), we can find all SFS values for this QAM scheme when $u=2$ MTs. 

In the multiple-MT ($u>2$) case, (\ref{eq:SFSt}) is no longer suitable for SFS values representation due to the increased number of MTs in MAC stage. In this case, the relationship between the values of SFSs and channel coefficients are no longer able to expressed by a simple ratio between different channel coefficients, e.g. the SFSs form different surfaces with infinite values in $3$-MT case. This is still an open issue in the literature for PNC design and we give our potential solution here. One solution is to utilise clashes instead of calculation of the values of SFSs in the proposed algorithm. In the multiple-MT case, an SFS still causes clashes and different clashes can be always found according to Definition $5$, then an optimal binary matrix is found if it maps the superimposed symbols within a clash to the same NCV and keep the value of $d_{min}$ maximised at the same time, without SFS calculation. 

Another solution to this issue is to divide the whole networks into multiple $2$-MT subnetworks. One way to achieve this goal is by allocating different pairs of MTs to different frequencies or time slots. In this case, the superimposed symbol at an AP is always from two MTs and then SFSs can be calculated by (\ref{eq:SFSt}). The only issue of this approach is that multiple On-line search algorithms for different MT pairs are required to be implemented which may cause extra computational complexity and latency time. An alternative way is to consider two MTs with the similar channel strength and transmit power as the prime MT pair and trade the other received signals as additional noise. According to our research in \cite{Acs}, when a strong signal with a much higher energy comapred to the rest of received signals is received at an AP in the multiple access stage, it is difficult to find an optimal matrix achieving unambigous recovery due to the high interference from this strong signals \cite{Acs}. Thus the $2$ MTs whose received signals are allocated at the similar energy level in the multiple access stage can be paired to form a subnetwork for PNC encoding and by pairing different MTs and APs, the multiple-MT-multiple-AP case is replaced by multiple $2$-MT-$2$-AP cases.

\subsection{Regulated BMAS Search Algorithm}
In order to fulfil the request of low latency in some scenarios, we present a regulated BMAS (R-BMAS) approach with a lookup table mechanism in this subsection. According to the definition of clash, the superimposed symbols in a clash have an intra-cluster distance of `$0$' and by the calculation in (\ref{eq:SFSt}), the clash groups in an SFS are mainly determined by the absolute value and the angle of the ratio of two channel coefficients. So following the design rules in the propose algorithm, we have:
\begin{theorem} \label{theorem:clsh}
The mapping matrix which resolves an SFS can always resolve the non-singular fade states with the values close to this SFS.
\end{theorem}
\begin{IEEEproof}
When a non-singular fade state (nSFS) happens, different superimposed symbols received at an AP will not be coincided which means no clash is observed. When this nSFS holds a similar absolute value and rotation angle to an SFS, the superimposed symbols, which form a clash in this SFS, will form a cluster in this nSFS with a smaller intra-cluster distances compared to inter-cluster distances to other clusters. In this case, the mapping matrices, that are capable to resolve the SFS by mapping the coincided superimposed symbols in a clash to the same NCV and keep different NCVs as far as possible, can achieve the maximum $d_{\mathrm{min}}$ in this nSFS by mapping the superimposed symbols in the cluster to the same NCV.
\end{IEEEproof}

An example of this theorem is given in Appendix \ref{ap:rBMAS}. According to Theorem 4, the proposed On-line search approach could be replaced by a lookup table based mechanism for the optimal mapping matrices selection. A table contains all SFS combinations and their corresponding invertible $mu \times mu$ optimal binary mapping matrices could be established in the Off-line search. During the real-time transmission, when the channel coefficients are estimated at the $j^{\mathrm{th}}$ AP, the value of the fade state $v_{FS_j}$ is calculated by
\begin{align} \label{eq:vFS}
v_{FS_j} = h_{j , 2} / h_{j , 1},
\end{align}
and then the closest SFS to this nSFS is obtained by
\begin{align}\label{eq:dFS}
& d^{(q_j)}_{FS} = \min|v_{FS_j} - v^{(q_j)}_{SFS}|^2, \\
\text{for} &~q = 1, \cdots, L, ~j = 1, \cdots, n.
\end{align}
The index $q_j$ will be forwarded to the CPU. By checking the table, the CPU send the optimal mapping matrix index back to the APs for PNC encoding. The algorithm is summarised in Algorithm \ref{Alg:regserch}.  


\begin{algorithm}[ht]
\caption{Regulated Binary Matrices Adaptive Selection (R-BMAS) Algorithm}
\label{Alg:regserch}
\begin{algorithmic}[1]
\Statex
\textbf{Off-line Search}
\For {$i=1:L$}  \Comment{each SFS}
\State {Apply Algorithm \ref{Alg:offline.p1} and \ref{Alg:offline.p2} for $\mathbf{M}_i$}
\EndFor
\For {$i_1=1:L$}  \Comment{all SFS combinations}
\State {\vdots} 
\For {$i_n=1:L$}  \Comment{all $n$ APs}
\State {$\mathbf{G} = \begin{bmatrix} \mathbf{M}_{i_1} \\
\vdots \\
\mathbf{M}_{i_n}
\end{bmatrix}$}
\State $\delta \leftarrow\mathrm{det}(\mathbf{G})_{|\mathbb{F}_2} $ \Comment{determinant over $\mathbb{F}_2$.}
\If{$\delta = 1$}
\State $\mathcal{G}_l\leftarrow \mathcal{G}_l\cup\mathbf{G}$ 
\State {Add $l$ as the optimal mapping matrix for SFS combination [$i_1\cdots i_n$]}
\EndIf
\EndFor
\State {\vdots}
\EndFor

\Statex
\textbf{On-line Search}
\For {$j=1:n$} \Comment{each AP}
\State {$v_{FS_j} = h_{2 , j} / h_{1 , j}, ~~j \in [1,2]$} \Comment{fade state}
\For {$i=1:L$}
\State {$\mathbf{d}^{(i_j)}_{SF} = |v_{FS_j} - v^{(i)}_{SFS}|^2$}
\EndFor
\State { $[d^{(k_j)}, k_j] = \min\mathbf{d}_{SF}^{(i_j)}$} \Comment{SFS index}
\EndFor

\State {Forward [$k_1\cdots k_n$] to CPU }
\State {Look up the table to obtain the optimal mapping matrix index $l$}
\State {Send the index back to APs}

\end{algorithmic}
\end{algorithm}


As shown in Algorithm \ref{Alg:regserch}, most of the calculations for mapping selection have been done before the transmission. At the same time, the latency is reduced by applying the regulated On-line search in Algorithm \ref{Alg:regserch} instead of Algorithm \ref{Alg:offline.p2}. However, a disadvantage of the R-BMAS algorithm is the performance degradation caused by sub-optimal global mapping matrices stored for some SFS combinations to achieve unambiguous recovery at the CPU. In order to overcome this problem, a combination of the two proposed algorithms can be utilised. During the Off-line search, a request of BMAS algorithm implementation could be stored in the table established in R-BMAS algorithm for the SFS combinations that need to be resolved by a suboptimal global mapping matrix. Thus the real-time computational complexity and latency time is reduced and the traffic in backhaul network is restricted to the total user data rate at the same time. 


\subsection{Computational Complexity and Backhaul Load}
We investigate the computational complexity of the ideal CoMP, non-ideal CoMP and the proposed algorithms in this subsection to illustrate the advantage of the proposed algorithms. 

In ideal CoMP, the bandwidth of backhaul network is assumed unlimited so that the received signal $y_j$ in (\ref{equ:channel}) at each AP will be forwarded to the CPU for joint multiuser ML detection, given by
\begin{align}
\mathbf{\hat{s}} = \arg\min\limits_{\mathbf{s}\in \Omega^u} \Vert \mathbf{y}-\mathbf{H}\mathbf{s}\Vert,
\end{align}
where $\mathbf{s}$ denotes the $u \times 1$ symbol vector at the MTs and $\mathbf{\hat{s}}$ is the estimated version, and $\mathbf{H}$ stands for the $n \times u$ channel matrix. 

In non-ideal CoMP, the backhaul network is bandwidth-limited and each AP employs an LLR based multiuser detection algorithm to estimate the transmitted symbols from each MT, mathematically given by
\begin{align}\label{eq:llr}
\chi_{i,\ell} = \frac{\sum\limits_{w_{i,\ell}=0}P(s_\ell)p(y_j\mid s_\ell)}{\sum\limits_{w_{i,\ell}=1}P(s_\ell)p(y_j\mid s_\ell)}, ~i=1,2,\cdots,m,
\end{align}
where $\chi_{i,\ell}$ stands for the LLR corresponding to the $i^\mathrm{th}$ bit of the binary message vector $\mathbf{w}_\ell$. A scalar quantizer which quantizes $\chi_{i,\ell}$ into binary bits is employed after the estimation. The quantised bits are sent to the CPU via the backhaul network. A detailed computation of (\ref{eq:llr}) is given in \cite{MUD}. Consider a quantisation scheme of $2$ bits is employed at each AP and $4$QAM is utilised at $2$ MTs in a simple $5$-node system, then $4$ LLRs are calculated by (\ref{eq:llr}) at each AP and a total $16$ bits are sent via the backhaul network. In order to achieve a good performance in terms of error rate and outage probability, a quantization scheme with a larger number of quantized bits is required and in this case, there is a trade-off between the performance and the backhaul load in non-ideal CoMP.  

In the proposed BMAS algorithm, each AP estimates the linear combination of the messages from MTs based on the ML rule rather than decoding individual symbols (such as ML detection in the ideal CoMP), mathematically given by (\ref{equ:PostProb.bg}) - (\ref{equ:PostProb.ed}). In order to minimise the computational complexity, an exhaustive search before the real-time transmission is implemented which contains the majority of computational complexity in the proposed algorithm, and the proposed On-line search is implemented during the transmission with a reduced number of mapping matrix candidates, e.g. $K=5$ matrices at each AP for $4$QAM modulation scheme in Algorithm \ref{Alg:offline.p2}.

In the proposed R-BMAS algorithm, calculations in the mapping selection are replaced by a lookup table mechanism and the computation complexity is reflected in distance comparison in (\ref{eq:dFS}). Then an LLR estimation of each bit in the NCV is applied which is the same as in the BMAS algorithm. Following the above $5$-node example, as illustrated in (\ref{eq:NCSt}), a binary mapping matrix with the minimum size of $2 \times 4$ is selected at each AP to encode the $4$ message bits from both MTs into the NCV, which results in a total $4$ bits backhaul load. Note an AP could employ a mapping matrix with the maximum size of $mu \times mu$ to generate the NCV and in this case, the other APs will not participate in the PNC encoding because they fail to receive any useful signals and the total backhaul load is still equal to $mu$.   

\section{Numerical Results}
\label{sec:NumRes}
In this section, we illustrate the outage probability performances of the proposed BMAS algorithm, R-BMAS algorithm and CoMP in a $5$-node system which includes $2$ MTs, $2$ APs and $1$ CPU. As mentioned in previous sections, the $5$-node network is the smallest network to apply the proposed algorithms so that we use this network as a  baseline to illustrate the advantage of the proposed algorithms. The proposed algorithms can be adapted to an N-MIMO network with more nodes and we have discussed the respective potential issues and solutions in Section V.  

In the simulations, we assume the multi-access links are wireless and the backhaul is wired which allows only binary data to be transmitted. Each node contains $1$ antenna for transmission and receiving, and $4$QAM/$16$QAM modulation schemes is employed at both MTs. We employ the convolutional code as an example and more powerful channel code can be utilised in order to enhance the reliability, such as LDPC \cite{Burr.MC}. In the simulation of ideal CoMP, we assume the backhaul capacity is limitless and the channel coefficients are exchanged in order to implement a joint ML detection algorithm. In the non-ideal CoMP scenario, quantizer with different quantization bits ($2$ bits and $4$ bits) are employed at each AP to investigate the outage probability performance.

\begin{figure}[ht]
  \centering
\begin{minipage}[t]{1\linewidth}
  \includegraphics[width=1.0\textwidth]{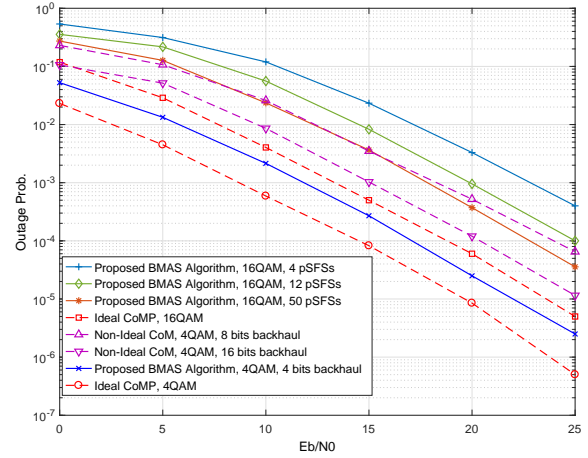} 
\end{minipage}
\caption{\small Outage Probability of the Proposed BMAS Algorithm in $4$QAM and $16$QAM.} \label{fig:SFS_16Q}
\end{figure}

Fig. \ref{fig:SFS_16Q} illustrates the outage probabilities of the proposed BMAS algorithm in $4$QAM and $16$QAM schemes. As we can see from the figure, the outage probability curves achieve the same diversity order. The ideal CoMP achieves the optimal performance in both $4$QAM and $16$QAM cases due to the unlimited backhaul capacity and joint detection. When the backhaul load is capacity limited, non-ideal CoMP with $2$-bit and $4$-bit quantizer, which results in a total of $8$ bits and $16$ bits backhaul load respectively, are implemented in the simulation. Compared to the ideal CoMP, a $8$dB and $13$dB performance degradation in the non-ideal CoMP can be observed; whilst the degradation is limited to only approximately $3$dB by the proposed algorithm. In the proposed BMAS algorithm, the backhaul load is equal to the total number of bits which is $4$ bits for $4$QAM, i.e. it is smaller than that in both non-ideal CoMP approaches. Instead of obtaining all SFSs-resolvable mapping matrix candidates for $16$QAM schemes, we consider only $4$, $12$ and $50$ pSFSs in Off-line search algorithm with computational complexity reduction. Note that the iSFSs are removed before selecting these $4$, $12$ and $50$ pSFSs in the simulation. As illustrated in the figure, approximately $10$dB degradation in outage performance is seen when using only $4$ pSFSs in the proposed On-line search algorithm. When $12$ and $50$ pSFSs are used, the degradation is reduced to $7$dB and $5$dB respectively and the gap will reduce when more pSFSs are considered in the proposed BMAS algorithm.

\begin{figure}[ht]
  \centering
\begin{minipage}[t]{1\linewidth}
  \includegraphics[width=1.0\textwidth]{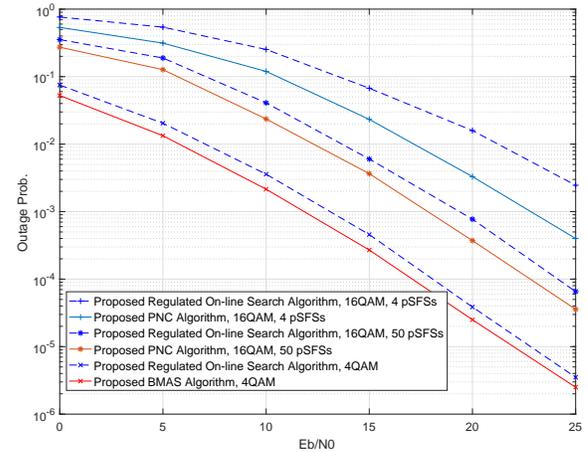} 
\end{minipage}
\caption{\small Outage Probability of the Proposed BMAS Algorithm vs Regulated On-line Search Algorithm.} \label{fig:regvsbmas}
\end{figure}

In Fig. \ref{fig:regvsbmas}, outage probability comparisons between the proposed BMAS algorithm and R-BMAS algorithm are shown. When $4$QAM is used at both MTs, outage probability performance of BMAS algorithm is $1$dB better than that of R-BMAS algorithm due to random suboptimal mapping matrices are stored in table used in the R-BMAS algorithm. In term of reduced computational complexity in the R-BMAS algorithm, an index searching among $25$ binary $4 \time 4$ mapping matrices are implemented to resolve all possible SFS combinations ($5$ pSFSs at each AP). When $16$QAM is employed at both MTs, the gap between the BMAS algorithm and the R-BMAS algorithm depends on the number of pSFSs utilised in the algorithm. When only $4$ pSFSs are used for optimal mapping matrix selection in both algorithms, BMAS achieves about $5$dB gains in outage performance compared to R-BMAS. The big gap in $4$ pSFSs case is caused by inefficient mapping matrix candidates in the R-BMAS algorithm. With the number of pSFSs increasing to $50$, more pSFS candidates are used in the R-BMAS algorithm which improves the outage performance about $7$dB and reduces the gap to $1$dB only.

\begin{figure}[t]
  \centering
\begin{minipage}[t]{1\linewidth}
  \includegraphics[width=1.0\textwidth]{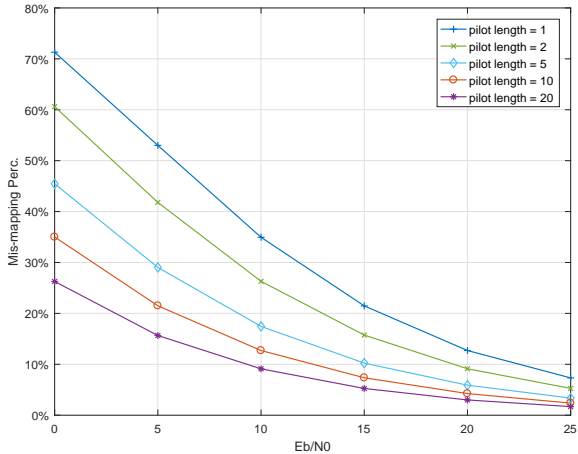} 
\end{minipage}
\caption{\small Miss-Mapping Probabilities with Different Pilot Lengths.} \label{fig:cha_map_prob}
\end{figure}

\begin{figure}[ht]
  \centering
\begin{minipage}[t]{1\linewidth}
  \includegraphics[width=1.0\textwidth]{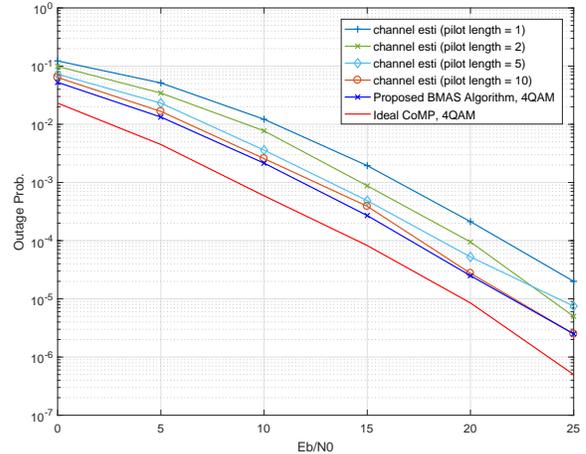} 
\end{minipage}
\caption{\small Outage Probability with Different Pilot Lengths.} \label{fig:SER_chaest}
\end{figure}

We have investigated how estimated channel state information (CSI) affects the network performance, and illustrated the result comparisons in Fig.s \ref{fig:cha_map_prob} and \ref{fig:SER_chaest}. The values of FSs are calculated by the first equation in (\ref{eq:SFSt}) so that during the transmission, the accuracy of the CSI in access link is important because it determines if the optimal mapping matrix can be selected. In Fig. \ref{fig:cha_map_prob}, we illustrate the impact of estimated CSI to the optimal mapping selection. The term \textquotedblleft mis-mapping \textquotedblright~ means the optimal mapping matrix is not selected. As we can see from the figure, the mis-mapping percentage decreases in any pilot length circumstances with increase of $E_b/N_0$. By using a short-length pilot sequence, the mis-mapping percentages are quite high which is caused by the fact that inaccurate fade states are calculated by using the estimated CSI. The low mis-mapping percentage shown in Fig. \ref{fig:cha_map_prob} leads to better outage probability performance in Fig. \ref{fig:SER_chaest}. For example, at $15$dB, the outage probability using only $1$ pilot symbol is $10^{-2}$ which refers to a mis-mapping percentage of $22\%$; while using $10$ pilot symbols, the outage probability reduces to $3 \times 10^{-3}$ and a mis-mapping percentage of only $8\%$ is achieved. By comparing the outage performances in Fig. \ref{fig:SER_chaest} with perfect and estimated CSI, we can conclude that pilot sequence with length of $10$ is good enough for the proposed BMAS algorithm.

\section{Future Work and Conclusion}
In this paper we present a design guideline of engineering applicable physical layer network coding in the uplink of N-MIMO networks. The proposed design criteria guarantee unambiguous recovery of all messages and the traffic in the backhaul network is reduced to the level of total user data rate at the same time. We then propose an optimal mapping matrix selection algorithm based on the design criteria. In order to reduce the real-time computational complexity, the proposed algorithm is divided into Off-line and On-line parts. An extension study of applying the proposed PNC design in binary systems with full-duplex (FD) APs \cite{FD-D2D}-\cite{FD-ARQ} has been started. Practical PNC design with cross layer optimisation in \cite{CLD-D2D}-\cite{CLD-D2D_2} provide another research direction, and the research in \cite{SE} focuses on spectrum efficiency solutions and can be extended to PNC-implemented systems. Moreover, the PNC application with optimal resource allocation \cite{RA-CDMA} and \cite{MQAM} is critical in order to serve the 5G systems and achieve massive data transmissions with high accuracy and low latency. The proposed algorithm is not only designed for a simple $5$-node network but for a general N-MIMO network serves multiple MTs. In addition, a regulated On-line search algorithm based on lookup table mechanism is also presented in order to further reduce the computational complexity and latency without much performance degradation. With reduced backhaul load, the proposed algorithms achieve higher outage probability performance compared to the practical non-ideal CoMP approaches.

\begin{appendices}

\section{}\label{ap:offline}
We illustrate a detailed Off-line search algorithm in Algorithms \ref{Alg:offline.p1} and \ref{Alg:offline.p2}. Part I indicates how to calculate SFSs and how to remove the image SFS in order to reduce the computational complexity, while Part II focuses on optimal mapping matrix selection. The steps in the proposed On-line search is the same as that in the Off-line search but with less number of matrix candidates so we will not show the repeated work here. $\mathcal{Q}_{\mathrm{d}}$ in Algorithm \ref{Alg:offline.p1} is defined as a vector contains all the $d_{min}$ between different NCSs for every binary mapping matrices in each SFS.
 
\begin{algorithm}[htb]
\caption{SFS Calculation and Image SFS Remove (Off-line Search Algorithm. Part I)}
\label{Alg:offline.p1}
\begin{algorithmic}[1]
\Statex
\For {$i=1:L$}  \Comment{each singular fade state}
\State {$h = \mathcal{S}(i)$} \Comment{$h$ is a $1\times m$ vector.}
\For{$j=1:K$} 					\Comment{each binary matrix}
\State $[\xi, T_{\xi}]= N(\mathcal{M}(j))$	
\State $\xi_{\mathrm{f}} \leftarrow \mathscr{F}(T_{\xi},h)$ \Comment{$\mathscr{F}(\cdot)$ produces all faded NCSs.}
\State $d_{\mathrm{min}} \leftarrow \mathscr{D}(\xi_{\mathrm{f}})$ \Comment{$\mathscr{D}(\cdot)$ calculates the minimum distance of all NCSs.}
\State $\mathcal{Q}_{\mathrm{d}}\leftarrow \mathcal{Q}_{\mathrm{d}} \cup  d_{\mathrm{min}}$ \Comment{store all $d_{\mathrm{min}}$ in $\mathcal{Q}_{\mathrm{d}}$.}
\EndFor
\State $[\boldsymbol\beta(i),\boldsymbol\alpha(i)] \leftarrow\mathscr{C}(\mathcal{Q}_{\mathrm{d}})$ \Comment{$\mathscr{C}(\cdot)$ sorts $\mathcal{Q}_{\mathrm{d}}$ in descending order stored in $\boldsymbol\beta(i)$ and outputs the rearranged index vector $\boldsymbol\alpha(i)$.}
\EndFor
\State $\mathcal{S}^{\prime} \leftarrow \mathscr{I}(\mathcal{S},\boldsymbol\alpha)$  \Comment{delete all image singular fade states and $\mathcal{S}^{\prime}$ has $L^{\prime}$ singular states, $L^{\prime}<L$.}
\State $\boldsymbol{\alpha}\leftarrow \boldsymbol\alpha \setminus \boldsymbol\alpha(\boldsymbol\beta=0)$ \Comment{delete the index element of $\boldsymbol\beta=0$.}
\State $\boldsymbol\alpha^{\prime} \leftarrow \boldsymbol{\alpha}(i|\mathcal{S}^{\prime})$ \Comment{$\boldsymbol\alpha^{\prime}$ corresponds to only  $\mathcal{S}^{\prime}$.}

\end{algorithmic}
\end{algorithm}


\begin{algorithm}[!h]
\caption{Binary Matrix candidates Selection for Each AP (Off-line Search Algorithm. Part II)}
\label{Alg:offline.p2}
\begin{algorithmic}[1]
\Statex
\For{$l_{L^{\prime}}=1:L^{\prime}$}
\State $\mathcal{S}^{\dag}_{L^{\prime}-1}\leftarrow\mathcal{S}^{\prime}\setminus \mathcal{S}^{\prime}(l_{L^{\prime}})$
\State $\theta_{L^{\prime}-1}\leftarrow \mathscr{E}(\mathcal{S}^{\dag}_{L^{\prime}-1})$ \Comment{Index set of $\mathcal{S}^{\prime}$ excluding the $l^{\mathrm{th}}$ element.}
\For{$l_{L^{\prime}-1}= \theta_{L^{\prime}-1}$ }
\State $\mathcal{S}^{\dag}_{L^{\prime}-2}\leftarrow\mathcal{S}^{\dag}_{L^{\prime}-1}\setminus \mathcal{S}^{\dag}_{L^{\prime}-1}(l_{L^{\prime}-1})$
\State $\theta_{L^{\prime}-2}\leftarrow \mathscr{E}(\mathcal{S}^{\dag}_{L^{\prime}-2})$
\State $\vdots$
\For{$l_{L^{\prime}-n+1} = \theta_{{L^{\prime}-n+1}}$}
\For{$i_1 = 1:K$} 
\State $\vdots$
\For{$i_n = 1:K$}
\State $
\mathbf{M} = \begin{bmatrix} \mathcal{M}[\boldsymbol\alpha(l_{L^{\prime}},i_1)] \\
\vdots \\
\mathcal{M}[\boldsymbol\alpha(l_{L^{\prime}-n+1},i_n)]
\end{bmatrix}
$
\State $\delta \leftarrow\mathrm{det}(\mathbf{M})_{|\mathbb{F}_2} $ \Comment{determinant over $\mathbb{F}_2$.}
\If{$\delta = 1$}
\State $\mathscr{R}\leftarrow \mathscr{R}\cup (l_{L^{\prime}}\cdots l_{L^{\prime}-n+1};i_1\cdots i_n)$
\State $\mathbf{G}\leftarrow \mathbf{G}\cup\mathbf{M}$ \Comment{$\mathbf{M} \leftrightarrow\mathbf{G}_{A(k)}$ in $\mathbf{G}$ has unique address $A(k)=(l_{L^{\prime}}^{(k)}\cdots l_{L^{\prime}-n+1}^{(k)};i_1^{(k)}\cdots i_n^{(k)})$, $k=1,\cdots \frac{L^{\prime}!}{(L^{\prime}-n)!}$.}
\State \Return {(21)}
\EndIf
\EndFor
\EndFor
\EndFor
\EndFor
\EndFor

\State $$[\mathbf{G}_{A(k_1)}\cdots\mathbf{G}_{A(k_n)}]\leftarrow\mathscr{X}(\mathbf{G})$$ \Comment{find $n$ $\mathbf{M}$ from $\mathbf{G}$ satisfying bijection relations $(l_{L^{\prime}}^{(k_e)}\cdots l_{L^{\prime}-n+1}^{(k_e)})\Leftrightarrow\mathcal{S}^{\prime}$ for $k_e = k_1\cdots k_n$.}
\For {$i=1:n$}
\State $\mathcal{Q}_i\leftarrow [\mathbf{G}_{A(k_1)}^{i}\cdots\mathbf{G}_{A(k_n)}^{i}]$ \Comment{$\mathbf{G}_{A(k_i)}^{i}=\mathcal{M}[\boldsymbol\alpha(l_{L^{\prime}-i+1}^{(k_i)},i_i^{(k_i)})]$}
\EndFor
\State \textbf{Output:} $n$ stacks $\mathcal{Q}_i$ with each including $L^{\prime}$ binary matrices.

\end{algorithmic}
\end{algorithm}


\section{}\label{ap:rBMAS}
\begin{figure}[ht]
  \centering
\begin{minipage}[t]{1\linewidth}
  \includegraphics[width=1.0\textwidth]{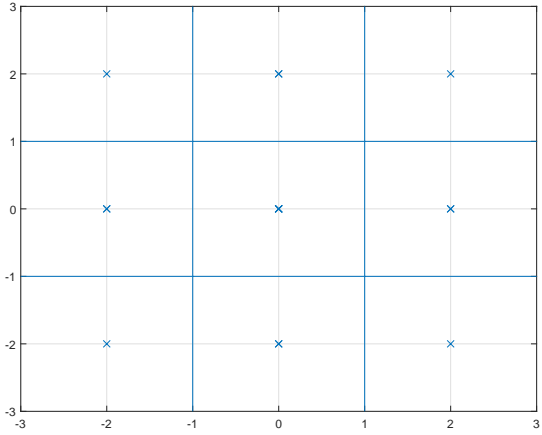} 
\end{minipage}
\caption{\small Constellation of the Received Signals at AP, $v_{SFS_1}=i$.} \label{fig:QAM4_SFS2}
\end{figure}

We illustrate an example of Theorem 4 here. In Fig. \ref{fig:QAM4_SFS2}, a received constellation of an SFS with  $v_{SFS_1}=i$ is illustrated. The number of MTs is $u=2$ and $4$QAM modulation is employed. We can observe the clashes clearly from the figure and their values can be calculated according to (\ref{eq:SFSt}), e.g. $4$ constellation points are superimposed at $(0,0)$ and $2$ are at $(0,2)$. Then the optimal binary mapping matrix will encode the superimposed constellation points in a clash to the same NCV and maximise the distance between different NCVs at the same time according to the design criteria.

\begin{figure}[ht]
  \centering
\begin{minipage}[t]{1\linewidth}
  \includegraphics[width=1.0\textwidth]{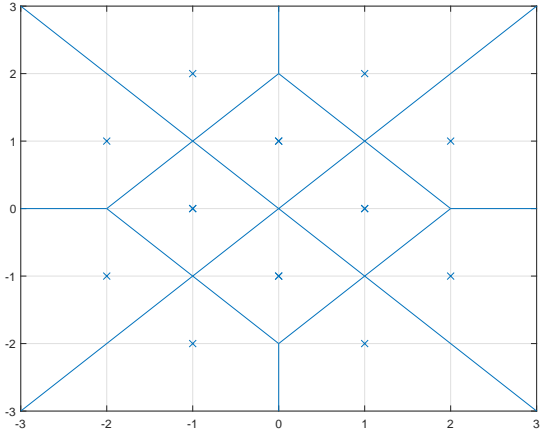} 
\end{minipage}
\caption{\small Constellation of the Received Signals at AP, $v_{SFS_1}=1/2+1/2i$.} \label{fig:QAM4_SFS3}
\end{figure}

Fig. \ref{fig:QAM4_SFS3} illustrates the received constellation of all possible superimposed symbols of another SFS with $v_{SFS_2}=1/2+1/2i$. In this case, $2$ constellation points are superimposed at $(0,1)$, $(0,-1)$, $(1,0)$ and $(-1,0)$, respectively. The optimal mapping matrices for $SFS_1$ and $SFS_2$ are different due to the differences between the clashed constellation points.

\begin{figure}[ht]
  \centering
\begin{minipage}[t]{1\linewidth}
  \includegraphics[width=1.0\textwidth]{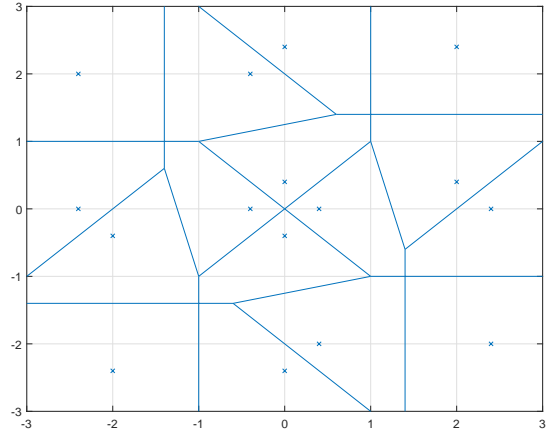} 
\end{minipage}
\caption{\small Constellation of the Received Signals at AP, $v_{nSFS}=7/10+7/10i$.} \label{fig:QAM4_SFS3_5}
\end{figure}

Then we consider a received constellation with a non-singular fading with $v_{nSFS}=7/10+7/10i$ illustrated in Fig. \ref{fig:QAM4_SFS3_5}. Clearly, none of the constellation points are superimposed but we can easily indicate different clusters by the distances between constellation points. Also we can find that the distance, which refers the absolute value and rotation angle in (\ref{dmint}), between $v_{SFS_1}$ and $v_{nSFS}$ is smaller than that between $v_{SFS_2}$ and $v_{nSFS}$. Then according to the unambiguous detection theorem, the $4$ points around $(0,0)$ in Fig. \ref{fig:QAM4_SFS3_5} should be mapped to the same NCV to maximise the inter-cluster distance. The same criteria should be satisfied by the $8$ points near $(0,2)$, $(-2,0)$, $(0,-2)$ and $(2,0)$. Then the initial cluster groups in Fig. \ref{fig:QAM4_SFS2} and Fig. \ref{fig:QAM4_SFS3_5} are the same which leads to the same optimal mapping matrices could be used in both circumstances. 
 
\end{appendices}

 \bibliographystyle{IEEEtran}
 {\footnotesize{
}}

\end{document}